\documentclass[10pt,twocolumn,letterpaper]{article}

\usepackage{iccv}
\usepackage{times}
\usepackage{epsfig}
\usepackage{graphicx}
\usepackage{amsmath}
\usepackage{amssymb}
\usepackage{pifont}
\usepackage{booktabs}
\usepackage{multirow}

\usepackage[breaklinks=true,bookmarks=false]{hyperref}

\iccvfinalcopy 

\def\iccvPaperID{****} 

\ificcvfinal\fi

\begin{document}

\title{SDWNet: A Straight Dilated Network with Wavelet Transformation for image Deblurring}

\author{Wenbin Zou$^{1, *}$, Mingchao Jiang$^{2, *}$, Yunchen Zhang$^{3, }$\thanks{Equal contribution}, Liang Chen$^{1, }$\thanks{Corresponding author}, Zhiyong Lu$^2$, Yi Wu$^1$\\
	Fujian Provincial Key Laboratory of Photonics Technology, Fujian Normal University, Fuzhou, China.$^1$\\
	JOYY AI GROUP, Guangzhou, China.$^2$\\
	China Design Group Co., Ltd., Nanjing, China.$^3$\\
	{\tt\small alexzou14@foxmail.com, jiangshaoyu1993@gmail.com, cydiachen@cydiachen.tech,}\\
	{\tt\small cl\_0827@126.com, yong1514@gmail.com, wuyi@fjnu.edu.cn}
	
	
}
\maketitle
\ificcvfinal\thispagestyle{empty}\fi

\begin{abstract}
	Image deblurring is a classical computer vision problem that aims to recover a sharp image from a blurred image. To solve this problem, existing methods apply the Encode-Decode architecture to design the complex networks to make a good performance. However, most of these methods use repeated up-sampling and down-sampling structures to expand the receptive field, which results in texture information loss during the sampling process and some of them design the multiple stages that lead to difficulties with convergence. Therefore, our model uses dilated convolution to enable the obtainment of the large receptive field with high spatial resolution. Through making full use of the different receptive fields, our method can achieve better performance. On this basis, we reduce the number of up-sampling and down-sampling and design a simple network structure. Besides, we propose a novel module using the wavelet transform, which effectively helps the network to recover clear high-frequency texture details. Qualitative and quantitative evaluations of real and synthetic datasets show that our deblurring method is comparable to existing algorithms in terms of performance with much lower training requirements. The source code and pre-trained models are available at \url{https://github.com/FlyEgle/SDWNet}.
\end{abstract}

\section{Introduction}

With the increasing ease of access to images, it is inevitable that blurred images will be obtained in different ways. It is increasingly important to eliminate the blur and restore a clear image. Since the process of image blurring is a one-to-many process, image deblurring is a notoriously difficult ill-posed problem in the field of image processing \cite{processing}. To address this problem, a number of optimization-based \cite{HighQuality,Forlearn,Anew,Unnatural,Blurkernel,dynamic,twophase} and learning-based methods \cite{Sun,MTRNN,Gong,Nah,Tao,Gao,Zhang,DeblurGAN,DeblurGAN-v2,Shen,Suin} have been proposed to learn the mapping function between the clear image and blurry image pairs. Most traditional deblurring methods \cite{Unnatural, Blurkernel, dynamic, twophase} tackle this problem via estimating blur kernel. Due to blur kernels in natural images are very complex, estimating the best blur kernel is a very tricky task. Therefore, inaccurate estimation of blur kernels results in poorly recovered images.

\begin{figure}
	\centering
	\includegraphics[scale=0.4,trim={6cm 0cm 8cm 0cm},clip]{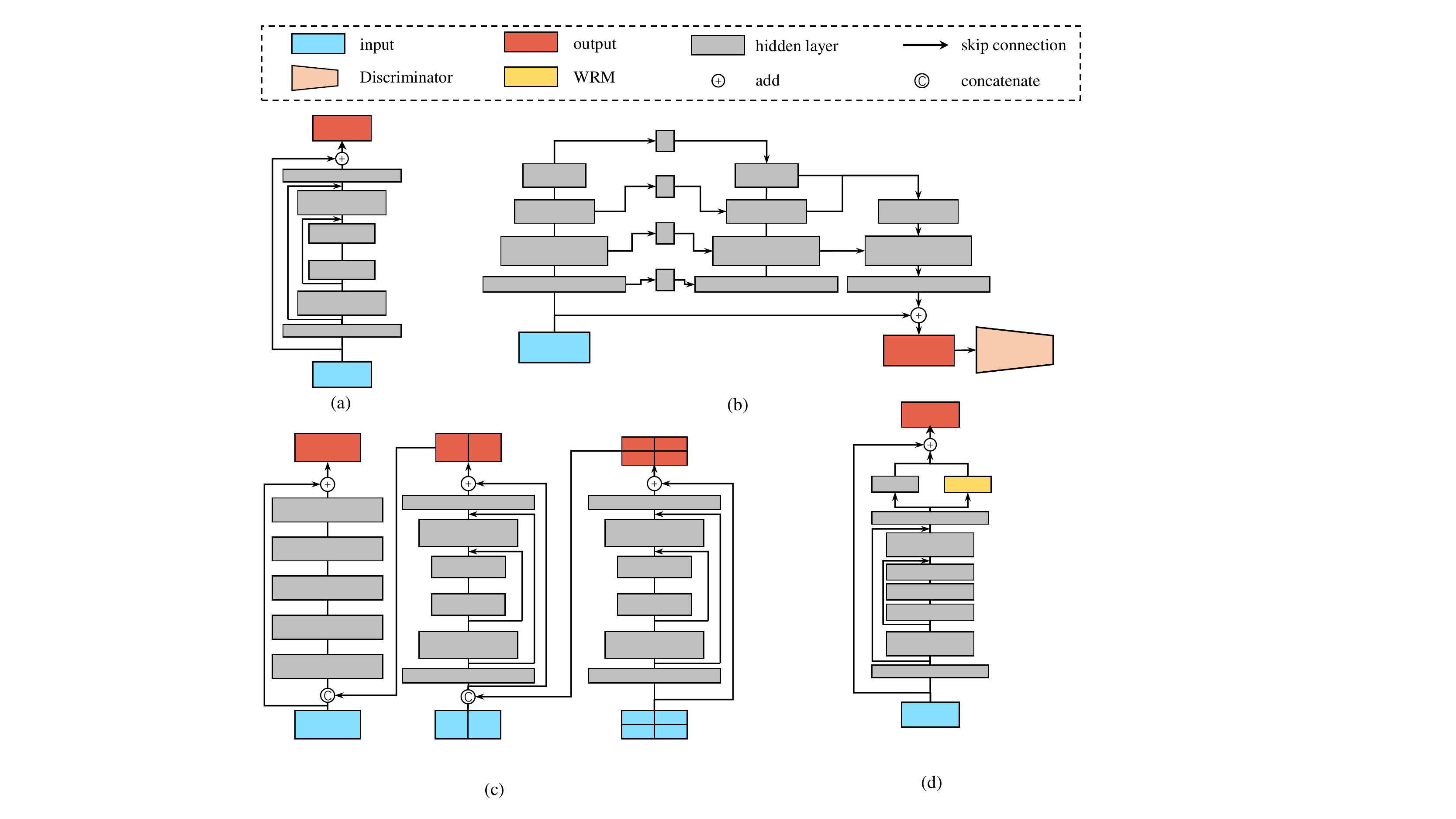}\\
	\caption{Schematic diagram of current mainstream network architecture. (a). Encode-Decode structure. (b). Generating adversarial network (GAN) structure (c). Coarse-to-fine structure. (d). Ours}
	\label{compare}
\end{figure}

Recently, the convolution neural network-based (CNN-based) algorithms achieve remarkable progress in image deblurring. Gong \textit{et al.} \cite{Gong} employ estimated dense motion flow maps to help the model learn the mapping between clear and blurred images. Then, Nah \textit{et al.}\cite{Nah} propose a multiscale loss function to implement a coarse-to-fine processing method and achieve good performance. However, this network is complex and very difficult to train. To address the difficulty of training, Tao \textit{et al.} \cite{Tao} and Gao \textit{et al.} \cite{Gao} improve the work by using shared network weights between different scales to achieve excellent performance. On this basis, Zhang \textit{et al.} \cite{Kzhang} propose an end-to-end CNN multilayer model similar to spatial pyramid matching. Kupyn \textit{et al.} \cite{DeblurGAN, DeblurGAN-v2} propose DeblurGAN and DeblurGAN-v2 based on adversarial learning to recover more realistic texture details from the blurry image. Shen \textit{et al.} \cite{Shen} propose a human-aware attentive deblurring network to remove the motion blur between foreground humans and background. Suin \textit{et al.} \cite{Suin} propose an efficient deblurring design built on new convolutional modules that learn the transformation of features using global attention and adaptive local filters to achieve superior performance. However, significant challenges remain in single image deblurring, as follows:
\begin{enumerate}
	\renewcommand\labelenumi{   \theenumi .   }
	\item Most of the above methods employ an Encode-Decode structure to learn the features of different receptive fields, as in Figure \ref{compare} (a). However, the repeated up-sampling and down-sampling contained in the Encode-Decode structure results in the loss of texture details, which affects the recovery of the image seriously.
	\item Some current image deblurring methods use GAN structures to obtain realistic texture details, as in Figure \ref{compare} (b). Since the GAN structure requires a joint generator and discriminator for training, it leads to unstable network performance.
	\item Most current image deblurring methods tend to design a coarse-to-fine structure to achieve superior PSNR performance, as shown in Figure \ref{compare} (c). However, coarse-to-fine structures are often very complex and computationally intensive resulting in a slow convergence process.
\end{enumerate}

In this paper, we address the above challenges using the method of dilated convolution and wavelet transform. We propose a novel image deblurring method that exploits the deblurring cues at different receptive filed via a dilated convolution model. Specifically, we propose a simple yet efficient end-to-end CNN model in the wavelet domain called straight dilated network with wavelet transformation (SDWNet), as in Figure \ref{compare} (d). It consists of the dilated convolution module and the wavelet reconstruction module. The dilated convolution module uses dilated convolution to obtain a larger field of perception for this network. This helps the model to capture similar features at a distance and thus facilitates image recovery. The wavelet reconstruction module provides additional information for spatial domain reconstruction by exploiting the frequency domain properties of the wavelet transform. Extensive experiments and ablation analysis demonstrate that with the assistance of the dilated convolution module and the wavelet reconstruction module, our SDWNet can achieve state-of-the-art performance.

The contributions of this work are summarized as follows:

\begin{itemize}
	\item We propose a dilated convolution module. Unlike previous deblurring networks that use repeated up-sampling and down-sampling to obtain large receptive fields, we use the dilated convolution with different dilated rates to obtain features with different receptive fields. This module facilitates the network to capture non-local similar features and recovers a clear image.
	\item We propose a wavelet reconstruction module. Instead of performing deblurring in a single spatial/frequency domain, we use the information recovered in the frequency domain to complement the spatial domain, so that the recovered image contains more high-frequency details.
	\item We propose a novel CNN-based image deblurring method. Different from previous deblurring methods that use a coarse to fine structure, we use a simple and streamlined structure to achieve results that are competitive with state-of-the-art methods. This structure effectively solves the problem of difficult training and slow convergence.
\end{itemize}

\begin{figure*}
	\centering
	\includegraphics[scale=0.3,trim={4cm 2cm 4cm 0.5cm},clip]{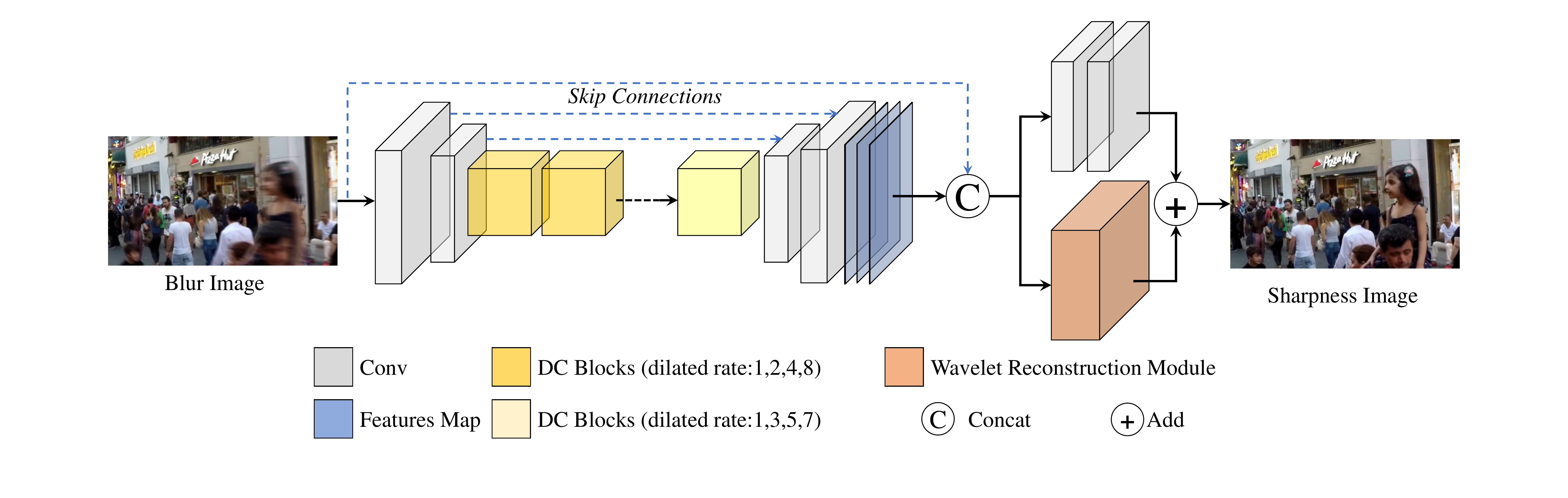}\\
	\caption{Network architecture of our proposed SDWNet.}
	\label{Network}
\end{figure*}

\section{Related Work}

\subsection{Deep Image Deblurring}
Recently, deep learning methods have achieved remarkable success in low-level computer vision tasks including image denoise \cite{VDCEDN}, image super-resolution \cite{EDSR,DWSR,MWCNN}, and image deblurring \cite{Sun,Nah,Tao,DeblurGAN,DeblurGAN-v2}. Many researchers tend to use deep learning methods to design an end-to-end model to achieve excellent performance. Sun \textit{et al.} \cite{Sun} design a CNN-based model to remove non-uniform motion blur by estimating the blur kernel. Due to the complexity of blurring in real scene images, the blur kernel estimation does not remove the fuzziness completely. Many deep learning-based methods tend to predict clear images directly from blurred images. Nah \textit{et al.} \cite{Nah} propose a multi-scale CNN model using a coarse-to-fine strategy, which can directly recover latent images without assuming any blur kernel. Because this network does not share parameters at different scales, it leads to increased computation and inference time. To address this problem, Tao \textit{et al.} \cite{Tao} propose an encoder-decoder structure with jump connections and parameter sharing at three scales, which effectively reduces computational effort and achieves better deblurring performance. Kupyn \textit{et al.} \cite{DeblurGAN, DeblurGAN-v2} propose DeblurGAN and DeblurGAN-v2 using adversarial learning and pyramidal structures to effectively recover clear images. Most of these networks perform alternate down-sampling and up-sampling of deep features to obtain large fields of perception. However, alternate up-sampling and down-sampling can cause a lot of information to be lost in the image recovery process, resulting in poor image recovery results. To address this problem, we use dilated convolution with different dilated rates to obtain information about the different receptive fields, thus making the recovered image clearer.

\subsection{Dilated Convolution}
Dilated convolution can obtain data features of different receptive fields by the jump step size and keeps the parameters constant. On this basis, dilated convolution has been successfully applied in many advanced vision tasks. Yu \textit{et al.} \cite{Yu} introduce dilated convolution for use on semantic segmentation, which significantly improves the segmentation performance. Zhou \textit{et al.} \cite{ZhouXiao} propose a cascade dilated module for medical image segmentation using convolutional layers with different dilated rates. Then, Brehm \textit{et al.} \cite{Brehm} introduce dilated convolution to the task of image deblurring and achieved excellent performance. They design an atrous convolution block using different dilated rates to recover sharper images. Due to the dilated rate of the atrous blocks follows the semantic segmentation method, the network does not completely cover all the pixel points, resulting in a still blurry image. We are inspired by their network and carefully adjust the dilated rate to obtain almost complete coverage of the receptive field.

\subsection{Related Application based on WT}
The wavelet transform is widely used in image processing tasks because it separates high-frequency information from low-frequency information in an image and is reversible. Many researchers introduce wavelet transforms into image restoration tasks \cite{Min, Zhang, MWCNN}. Min \textit{et al.} \cite{Min} use the wavelet transform to separate the frequency information from the blurred image and then recover the image, effectively weakening the smoothing characteristics of the image. Zhang \textit{et al.} \cite{Zhang} propose double discrete wavelet transform to enhance the blurred image processing capabilities. Liu \textit{et al.} \cite{MWCNN} propose a multilayer wavelet CNN using the U-Net structure, resulting in a clearer recovered image. These methods all use a direct mix of all frequency information, leading to problems with different frequency information interacting with each other and creating wrong textures. Therefore, we propose a wavelet transform reconstruction module that effectively recovers a clear image.

\section{Proposed Method}

\subsection{Framework}
In this section, we describe our proposed straight dilated network with wavelet transformation in detail. Since complex models can bring problems such as unstable training and slow convergence, thus we used a plain network structure, as shown in Figure \ref{Network}. Our SDWNet mainly consists of three parts: the shallow feature extraction layer, the dilated convolution (DC) module, and the reconstruction module. To obtain a larger perceptual field, we first utilize a kernel size of $7\times7$ convolution to extract shallow features. Inspired by the \cite{ZhouXiao}, we propose the dilated convolution blocks for fusing multi-receptive field information by using different dilated rates. Then, our network uses cascading multiple DC blocks to learn the broad contextual information. Due to the wavelet transform is an effective tool for recovering high-frequency information, we propose a wavelet reconstruction module as a parallel reconstruction branch, thereby preserving the desired fine texture in the final output image.

Unlike the cascade of multiple dilated convolution blocks with the same dilated rate in \cite{ZhouXiao}, we designed two dilated convolution blocks with different dilated rates to obtain richer receptive field information. Besides, we add jump connections to make full use of the information from the shallow features. Instead of other wavelet transform methods that predict the four frequency subbands directly, our method uses a shared network to recover the four frequency subbands separately, thus effectively avoiding artifacts caused by the interaction of different frequency subbands.

Given an input blurred image $\textbf{I}_{blur}$, the proposed model predicts a residual image $\textbf{R}$ to which the degraded input image $\textbf{I}_{blur}$ is added to obtain: $\textbf{X}=\textbf{I}_{blur}+\textbf{R}$. We optimize our SDWNet with the following loss function:
\begin{equation}
\mathcal{L}_{total} = \mathcal{L}_{char}(\textbf{X}, \textbf{Y}) + \lambda *\mathcal{L}_{ssim}(\textbf{X}, \textbf{Y}),\label{con:loss}
\end{equation}
where $\textbf{Y}$ represents the ground-truth image, and $\mathcal{L}_{char}$ is the Charbonnier loss \cite{charloss}:
\begin{equation}
\mathcal{L}_{char}=\dfrac{1}{N}\sum_{i=1}^{N}\sqrt{||\textbf{X}^i-\textbf{Y}^i||^2+\epsilon^2},
\end{equation}
with constant $\epsilon$ emiprically set to $10^{-3}$ for all the experiments. In addition, $\mathcal{L}_{ssim}$ is the ssim loss, defined as:
\begin{equation}
\mathcal{L}_{ssim} = \dfrac{1}{N}\sum_{i=1}^{N} SSIM(\textbf{X}^{i},\textbf{Y}^i),
\end{equation}
where $SSIM(\cdot)$ denotes the SSIM \cite{SSIM} operator. The parameter $\lambda$ in Eq. (\ref{con:loss}) is a hyper-parameter used to control the composition of the SSIM loss function. The following experiments will verify it. Next, we describe each key element of our method.

\begin{figure}
	\centering
	\includegraphics[scale=0.3,trim={1.2cm 0cm 3cm 0cm},clip]{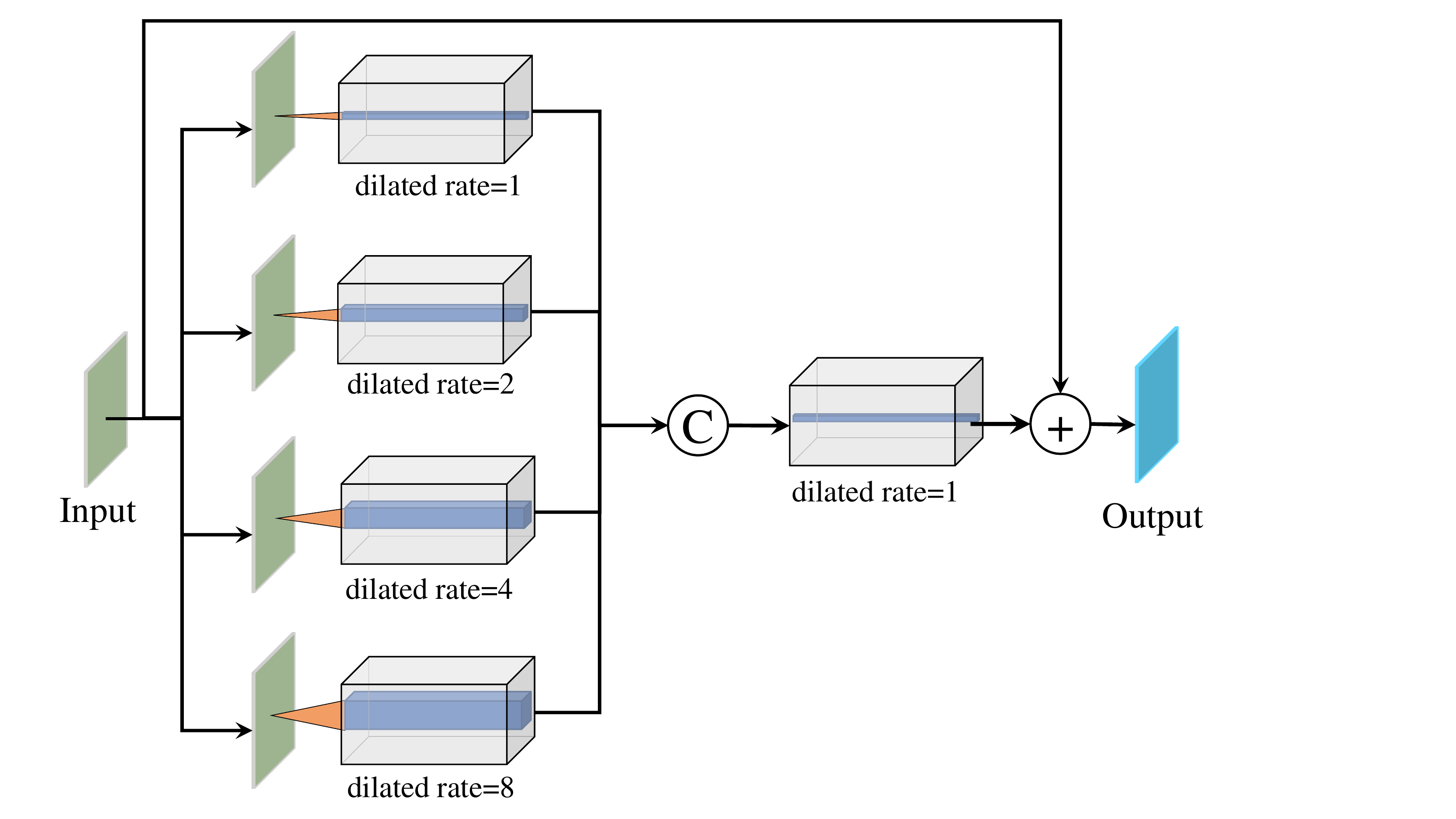}\\
	\caption{The architecture of our proposed dilated convolution block (DCB). We control the output of the different receptive fields by modifying the dilated rate of the intermediate parallel dilated convolution. We set the dilated rate to \{1, 3, 5, 7\} in the last layer of DCB. All other DCB dilated rates are set to \{1, 2, 4, 8\}.}
	\label{DCB}
\end{figure}

\subsection{Dilated Convolution Module}
We now give more details about our proposed dilated convolution module, which contains $n$ dilated convolution blocks (DCB). The DC Module is formulated as:
\begin{equation}
F_n = H_{DCB}^{n}(H_{DCB}^{n-1}(\cdots H_{DCB}^{1}(F_0)\cdots)),
\end{equation}
where $H_{DCB}^n$ denotes the function of $n$-th DCB. $F_n$ and $F_{1}$ represent the input and output of the DC Module. DCB is composed of multiple dilatied convolutions with different dilated rates in parallel, as shown in Figure \ref{DCB}. It can be expressed as follows:
\begin{eqnarray}
&F_{dr\_1} = H_{dr\_1}(F_{input}),\\
&F_{dr\_2} = H_{dr\_2}(F_{input}),\\
&F_{dr\_4} = H_{dr\_4}(F_{input}),\\
&F_{dr\_8} = H_{dr\_8}(F_{input}),\\
&F_{dr\_cat} = Concat(F_{dr\_1}, F_{dr\_2}, F_{dr\_4}, F_{dr\_8}),
\end{eqnarray}
where $H_{dr\_1}$, $H_{dr\_2}$, $H_{dr\_4}$, and $H_{dr\_8}$ denote dilated convolution operations with dilated rates of 1, 2, 4 and 8, respectively. $F_{dr\_1}$, $F_{dr\_2}$, $F_{dr\_4}$, and $F_{dr\_8}$ denote the output of dilated convolutions with different dilated rates. Inspired by \cite{dilation}, we attach fine-grained control on receptive fields. On shallow layers, we adopt regular dilated rates of 1, 2, 4, and 8. On the last layer, we adopt a non-overlapped dilated rate of 1, 3, 5, and 7 to avoid gridding effects for the image deblurring tasks. Then, we use a dilated convolution with a dilated rate of 1 to fuse features from different receptive fields. Finally, we superimpose the fused features onto the input features to get the output. The output features can be written as:
\begin{eqnarray}
&F_{fuse} = H_{fuse}(F_{dr\_cat}),\\
&F_{out} = F_{input} + F_{fuse},
\end{eqnarray}
where $H_{fuse}$ denotes the dilated convolution used to fuse the features. $F_{fuse}$ and $F_{out}$ denote the fused features and output features. 

\begin{figure}
	\centering
	\includegraphics[scale=0.27,trim={1.5cm 0cm 0cm 3cm},clip]{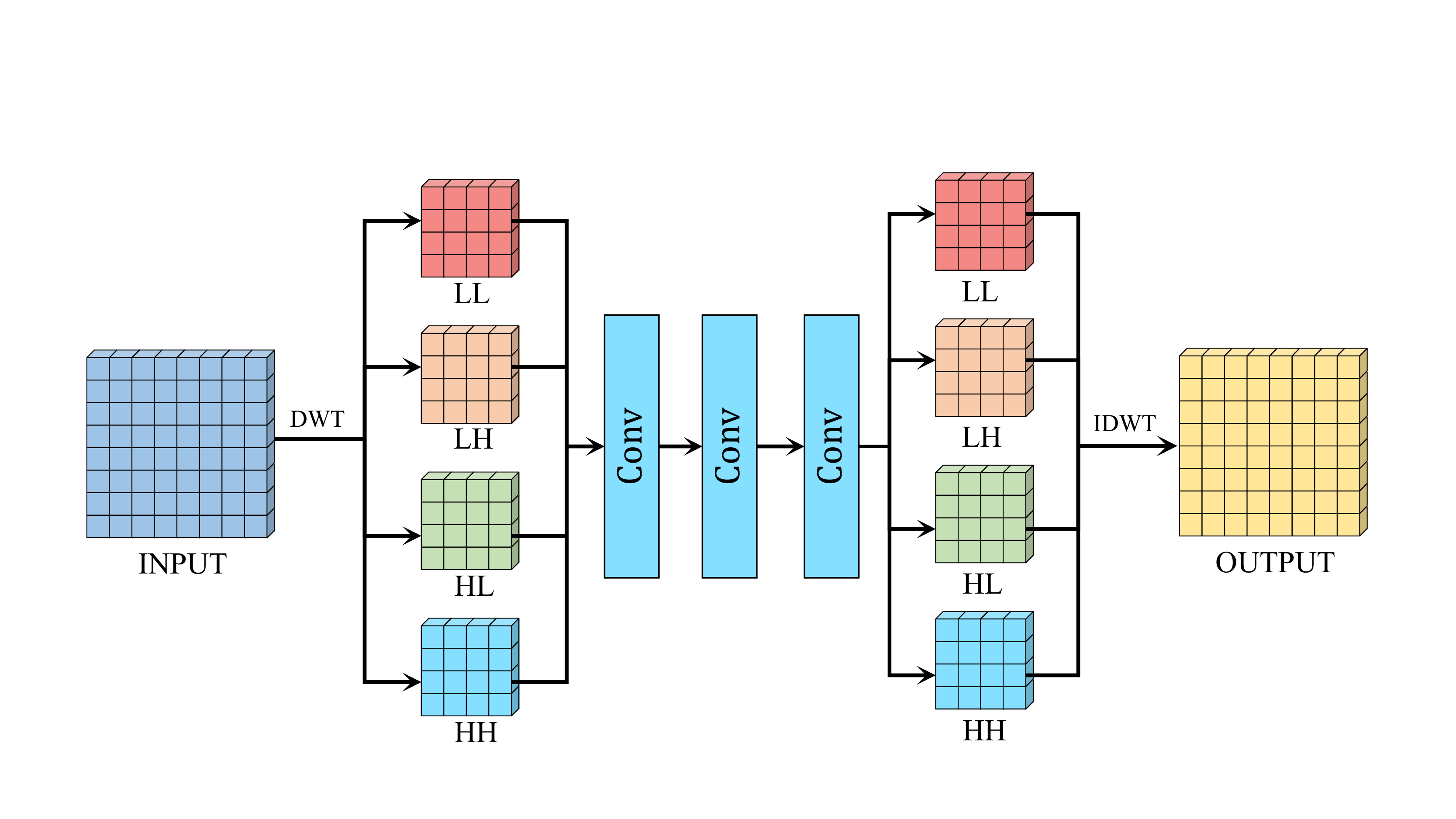}\\
	\caption{The architecture of our proposed wavelet reconstruction module (WRM). We decompose the input features into four frequency subbands by the wavelet transforms: LL, LH, HL and HH.Then, the corresponding frequency sub-bands are recovered by three-layer convolution. The final output is obtained using the wavelet inverse transform.}
	\label{WRM}
\end{figure}

\subsection{Wavelet Reconstruction Module}
The wavelet reconstruction module (WRM) mainly uses the wavelet transform to convert spatial domain information to the wavelet domain for recovery. As shown in Figure \ref{WRM}, the input feature $F_{input}$ can be divided into four different frequency sub-bands by the discrete wavelet transform. These frequency sub-bands can be defined as follows:
\begin{eqnarray}
\{F_{LL}, F_{LH}, F_{HL}, F_{HH}\} = \textbf{DWT}(F_{input}),
\end{eqnarray}
where $\textbf{DWT}(\cdot)$ denotes the operation of the discrete wavelet transform. $F_{LL}, F_{LH}, F_{HL}$, and $F_{HH}$ denote the feature of four frequency sub-bands, respectively. To avoid interference between the different frequency subbands, each of the four subbands is fed into a 3-layer convolutional network for recovery. We can express it as:
\begin{eqnarray}
&F_{LL\_r} = H_{conv\times 3}(F_{LL}),\\
&F_{LH\_r} = H_{conv\times 3}(F_{LH}),\\
&F_{HL\_r} = H_{conv\times 3}(F_{HL}),\\
&F_{HH\_r} = H_{conv\times 3}(F_{HH}),
\end{eqnarray}
where $H_{conv\times 3}(\cdot)$ denotes the 3-layer convolution network. $F_{LL\_r}$, $F_{LH\_r}$, $F_{HL\_r}$, and $F_{HH\_r}$ represent the four frequency sub-band features recovered by the 3-layer convolution network. We finally use the discrete wavelet inverse transform to reconstruct the recovered frequency sub-bands into output features $F_{out}$. It can be formulated as:
\begin{eqnarray}
F_{out} = \textbf{IDWT}(F_{LL\_r}, F_{LH\_r}, F_{HL\_r}, F_{HH\_r}),
\end{eqnarray}
where $\textbf{IDWT}(\cdot)$ denotes the discrete wavelet inverse transform operation. 

\section{Experiments with Analysis}
\subsection{DataSet}
The following are the training and test sets that we use:

The \textbf{GoPro} Dataset \cite{Nah} uses the GoPro Hero 4 camera to capture 240 frames per second (fps) video sequences, and generates blurred images through averaging consecutive short-exposure frames. It is a common benchmark for image motion blurring, containing 3,214 blurry/clear image pairs. We follow the same split \cite{Nah}, to use 2,103 pairs for training and the remaining 1,111 pairs for evaluation.

The \textbf{HIDE} Dataset \cite{Shen} is specifically collected for human-aware motion deblurring and its test set contains 2,025 images. While the GoPro and HIDE datasets are synthetically generated, the image pairs of the RealBlur dataset are captured in real-world conditions.

The \textbf{RealBlur} dataset \cite{realblur} has two subsets: (1). RealBlur-J is formed with the camera JPEG outputs, and (2). RealBlur-R is generated offline by applying white balance, demosaicking, and denoising operations to the RAW images.

\begin{figure}
	\centering
	\includegraphics[scale=0.38,trim={5.5cm 5.8cm 6.5cm 5.5cm},clip]{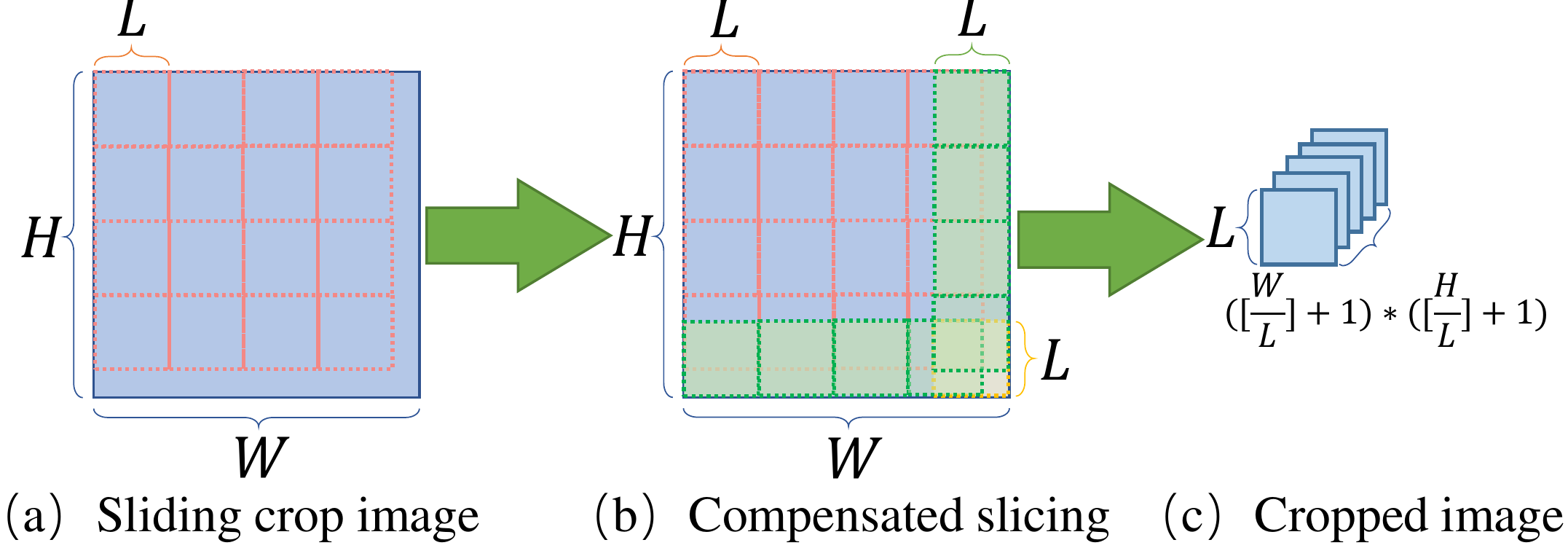}\\
	\caption{Dataset sliding crop, where $[\cdot]$ denotes represents the rounding down operation. (a) represents the cropping of a block of size $L\times L$ from an image of species $H\times W$. (b) indicates compensatory cropping at the edges of images not covered in (a), marked as green and yellow blocks. The result is (c) a series of the cropped image.}
	\label{fig3}
\end{figure}

\subsection{Dataset Sliding Crop}
To further improve the robustness of the network, we perform appropriate sliding window slicing on GoPro, as shown in Figure \ref{fig3}. The GoPro dataset images are all $1280\times720$ resolution, so we use a step of 240 to perform $480\times480$ size sliding window slicing in the order of left, right, top and bottom, and compensated slicing on the edge part. Finally, we can crop out 24 patches from each image. Thus, we can crop up to 50472 patches from the original dataset.

\subsection{Implementation Details}
We implement all of the models using PyTorch \cite{pytorch}. Our SDWNet is an end-to-end trainable network and requires no pretraining. Compare with others methods, our network needs fewer training epochs. In the training stage, we use the AdamW \cite{AdamW} optimizer to train our model. We set the input size to $416\times 416$ and the batch size to 8. The initial learning rate of $4\times10^{-4}$, and we use the Cosine Annealing strategy \cite{SGDR} to steadily decrease the learning rate. Weights decay is setting as $1\times10^{-4}$ for the regularization model. We set the hyperparameter $\lambda$ in the loss function to 1. And, we use a data augmentation strategy of random rotation, random flip, and RGB channel shuffle. We first use the GoPro datasets to train 1500 epochs with the above configuration. Then, we train 50 epochs on the GoPro crop datasets with the best model to get the best results. Besides, all experiments are conducted on the desktop computer with two NVIDIA Tesla V100 GPUs.

\begin{table}
	\caption{Quantitative comparisons of our models with the state-of-the-art deblurring methods on GoPro \cite{Nah} and HIDE \cite{Shen} datasets (PSNR(dB)/SSIM). Best and second-best results are \textbf{highlighted} and \underline{underlined}. $^\circledast$ represents the training results of our method on a cropped GoPro dataset.}
	\centering
	\resizebox{8cm}{!}{
		\begin{tabular}{l|c|c|c|c}
			\toprule[1.2 pt]
			\multirow{2}{*}{Method}             &      \multicolumn{2}{c|}{GoPro}       &       \multicolumn{2}{c}{HIDE}        \\[1pt]
			& {PSNR$\uparrow$}  & {SSIM$\uparrow$}  & {PSNR$\uparrow$}  & {SSIM$\uparrow$}  \\[1pt] \hline
			Xu \textit{et al.} \cite{Unnatural} &       21.00       &       0.741       &         -         &         -         \\[1pt]
			Hyun \textit{et al.} \cite{dynamic} &       23.64       &       0.824       &         -         &         -         \\[1pt]
			Whyte \textit{et al.} \cite{whyte}  &       24.60       &       0.846       &         -         &         -         \\[1pt]
			Gong \textit{et al.} \cite{Gong}    &       26.40       &       0.863       &         -         &         -         \\[1pt]
			DeblurGAN \cite{DeblurGAN}          &       28.70       &       0.858       &       24.51       &       0.871       \\[1pt]
			Nah \textit{et al.} \cite{Nah}      &       29.08       &       0.914       &       25.73       &       0.874       \\[1pt]
			Zhang \textit{et al.} \cite{PZhang} &       29.19       &       0.931       &         -         &         -         \\[1pt]
			DeblurGAN-v2 \cite{DeblurGAN-v2}    &       29.55       &       0.934       &       26.61       &       0.875       \\[1pt]
			SRN  \cite{Tao}                     &       30.26       &       0.934       &       28.36       &       0.915       \\[1pt]
			Shen \textit{et al.} \cite{Shen}    &         -         &         -         &       28.89       &       0.930       \\[1pt]
			Gao \textit{et al.} \cite{Gao}      &       30.90       &       0.935       &       29.07       &       0.913       \\[1pt]
			DBGAN \cite{DBGAN}                  &       31.10       &       0.942       &       28.94       &       0.915       \\[1pt]
			MT-RNN \cite{MTRNN}                 &       31.15       &       0.945       &  \underline{29.15}   &      {0.918}      \\[1pt]
			DMPHN \cite{DMPHN}                  &       31.20       &       0.940       &       29.09       &       0.924       \\[1pt] \hline
			\textbf{SDWNet(Ours)}                        & \underline{31.26} & \underline{0.966} &       28.99       & \underline{0.957} \\[1pt]
			\textbf{SDWNet$^\circledast$(Ours)}                &  \textbf{31.36}   &  \textbf{ 0.967}  & \textbf{29.23} &  \textbf{0.963}   \\[1pt] \bottomrule[1.2 pt]
		\end{tabular}
	}
	\label{Gopro}
\end{table}

\begin{table}
	\caption{Deblurring comparisons on the RealBlur dataset \cite{realblur} under two different settings: 1). applying our GoPro trained model directly on the RealBlur set (to evaluate generalization to real images), 2). Training and testing on RealBlur data where methods are denoted with symbol $^\star$. The PSNR/SSIM scores for other evaluated approaches are taken from the RealBlur benchmark \cite{realblur}. }
	\centering	
	\resizebox{8cm}{!}{
		\begin{tabular}{l|c|c|c|c}
			\toprule [1.2 pt]
			\multirow{2}{*}{Method}             &    \multicolumn{2}{c|}{RealBlur-R}    &    \multicolumn{2}{c}{RealBlur-J}     \\[1pt]
			& {PSNR$\uparrow$}  & {SSIM$\uparrow$}  & {PSNR$\uparrow$}  & {SSIM$\uparrow$}  \\[1pt] \hline
			Hu \textit{et al.} \cite{Hu}        &       33.67       &       0.916       &       26.41       &       0.803       \\[1pt]
			Nah \textit{et al.} \cite{Nah}      &       32.51       &       0.841       &       27.87       &       0.827       \\[1pt]
			DeblurGAN \cite{DeblurGAN}          &       33.79       &       0.903       &       27.97       &       0.834       \\[1pt]
			Pan \textit{et al.} \cite{Pan}      &       34.01       &       0.916       &       27.22       &       0.790       \\[1pt]
			Xu \textit{et al.} \cite{Unnatural} &       34.46       &       0.937       &       27.14       &       0.830       \\[1pt]
			DeblurGAN-v2 \cite{DeblurGAN-v2}    &       {35.26}       &       0.944       &  \textbf{28.70}   &      {0.866}      \\[1pt]
			Zhang \textit{et al.} \cite{PZhang} &       35.48       &       0.947       &       27.80       &       0.847       \\[1pt] 
			SRN                    \cite{Tao}   &       35.66       &       0.947       &       28.56       &       \underline{0.867}       \\[1pt] 
			DMPHN                \cite{DMPHN}   &       \underline{35.70}       &       \underline{0.948}       &       {28.42}       &       {0.860}       \\[1pt] 
			\textbf{SDWNet(Ours)}                        & \textbf{35.85} & \textbf{0.948} &       \underline{28.61}       & \textbf{0.867} \\[1pt]\hline\hline
			DeblurGAN-v2$^\star$ \cite{DeblurGAN-v2}    &       {36.44}       &       0.935       &  {29.69}   &      {0.870} \\[1pt]
			\textbf{SDWNet$^\star$(Ours)}                &  \textbf{38.21}   &  \textbf{0.963}   & \textbf{30.73} &  \textbf{0.896}   \\[1pt] \bottomrule[1.2 pt]
		\end{tabular}
	}
	\label{realblur}
\end{table}

\begin{figure*}
	\centering
	\includegraphics[scale=0.12,trim={3cm 12cm 3cm 16cm},clip]{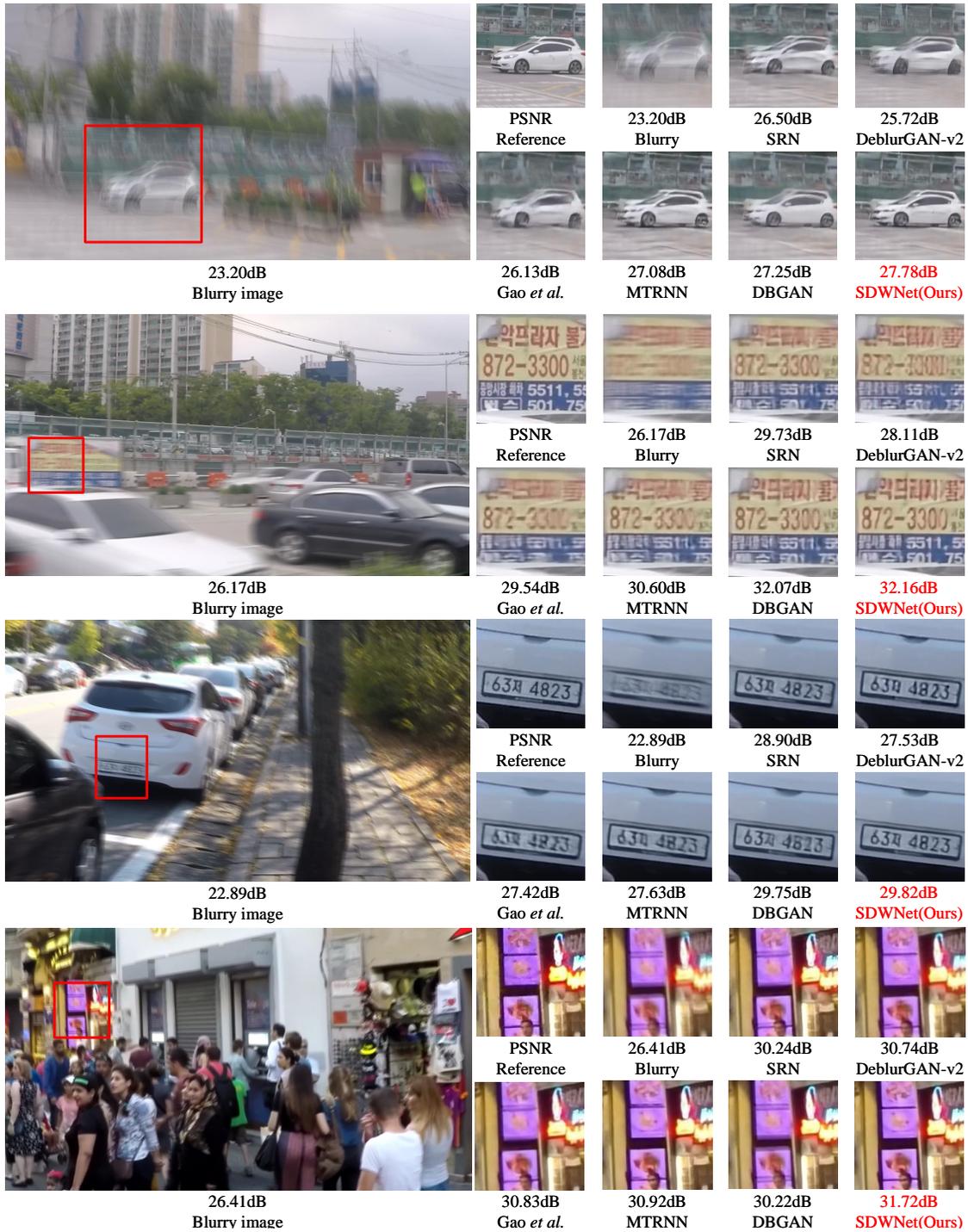}\\
	\caption{Qualitative comparison with the leading deblurring algorithms: SRN \cite{Tao}, DeblurGAN-v2 \cite{DeblurGAN-v2}, Gao \textit{et al.} \cite{Gao}, MTRNN \cite{MTRNN}, and DBGAN \cite{DBGAN}. From the figure, we can see that our method can generate the right and clear details of the image.}
	\label{cpimage}
\end{figure*}

\subsection{Image Deblurring Results}
\textbf{Quantitative results.}
Quantitative analyses are performed to evaluate the performance of the SDWNet for image deblurring. More precisely, we quantitatively assess the average performance of PSNR and SSIM over GoPro and HIDE datasets. We compare our SDWNet with the excellent deblurring methods \cite{Unnatural,dynamic,whyte,Gong,DeblurGAN,Nah,PZhang,DeblurGAN-v2,MTRNN,DMPHN} of the past and the experimental results are shown in Table \ref{Gopro}. From Table \ref{Gopro} it can be seen that our method can achieve better performance compared with other deblurring methods. Compared with the previous DMPHN method, our method achieves 0.16dB improvement in PSNR and 0.027 improvements in SSIM. It is worth noting that not only does our method achieve the best performance on the GoPro dataset, but it also achieves the best results on the HIDE at the same time. 

To demonstrate the generalization performance of our method in real scenarios, we also perform experimental validation on the RealBlur dataset, as shown in Table \ref{realblur}. Compared to previous best deblurring methods, our SDWNet achieves the best performance on the RealBlur-R dataset. Our method achieves a 2.51dB PSNR performance gain on the RealBlur-R dataset. On the RealBlur-J dataset, we obtain similar PSNR and SSIM performance to the previous best methods.

\textbf{Visual results.}
A qualitative analysis of the effect of our SDWNet on image deblurring compared with other methods is shown in Figure \ref{cpimage}. We compare the visual deblurring result of our method with the previous methods. To fully demonstrate the superiority of our method, we have zoomed in on the details in the image shown. It is worth noting that many of the detailed textures in the blurred image are difficult to determine. As a result, the repeated upsampling and downsampling process can cause the texture orientation to change, which can affect image performance. We solve this problem effectively by the dilated convolution to enable the recovered details to be correct. Our method also uses the wavelet transform to convert the features to the frequency domain for recovery, ensuring full recovery of high-frequency details. Therefore, our approach visual results in superior performance. Using the second image in Figure \ref{cpimage} as an example, the images recovered by the older methods still have some blurring. The more recent methods of the last few years have produced images with some error texture. However, our proposed SDWNet can accurately recover a clear image. It demonstrates that our method outperforms other methods in qualitative analysis.

\textbf{Performance and efficiency comparison.} In addition to the superior PSNR and SSIM of our model, we also compare the parameters and running times of our method with the previous methods. The results of the experiment are shown in Table \ref{runtime}. Our method has competitive PSNR and SSIM performance to other superior methods, but the parameters and FLOPs of our method are much smaller than other methods, and our method is the fastest in Table \ref{runtime}. Notably, our method achieves better PSNR and SSIM performance than DMPHN \cite{DMPHN} using only one-third of the parameters and FLOPs of DMPHN \cite{DMPHN}. It efficiently demonstrates that the efficient deblurring performance of our network structure.

\begin{table*}
	\caption{Performance and efficiency comparison on the GoPro \cite{Nah} test dataset. Runtimes are computed with the Nvidia Titan Xp GPU. FLOPs are computed with the input size of $256\times256$.}
	\centering
	\resizebox{13cm}{!}{
		\begin{tabular}{l|c|c|c|c|c|c}
			\toprule[1.2 pt]
			Method     & DeblurGAN-v2 \cite{DeblurGAN-v2} & DBGAN \cite{DBGAN} & DMPHN \cite{DMPHN} & Suin \textit{et al.} \cite{Suin} & MPRNet \cite{MPRNet}  & SDWNet (Ours)   \\[1pt] \hline\hline
			Params (M) &               60.9               &        11.6        &        21.7        &               23.0               &         20.1      & \textbf{7.2}    \\[1pt]
			Flops (G)  &              411.34              &       660.2        &       678.56       &              536.74              &        660.2      & \textbf{181.31} \\[1pt]
			Time (s)   &               0.21               &        0.83        &        1.07        &               0.34               &         0.18        & \textbf{0.14}   \\[1pt]
			PSNR       &              29.55               &       31.10        &       31.20        &              31.85               &    \textbf{32.66}    & 31.36           \\[1pt]
			SSIM       &              0.934               &       0.942        &       0.940        &              0.948               &        0.959         & \textbf{0.967}  \\[1pt] \bottomrule[1.2 pt]
		\end{tabular}
	}
	\label{runtime}
\end{table*}

\begin{table}
	\caption{Model Policy with depths and widths on network performance with an input of $96\times 96$}
	\centering
	\resizebox{5cm}{!}{
		\begin{tabular}{l|c|c}
			\toprule[1.2 pt]
			Model Setting  &      PSNR      &      SSIM       \\[1pt] \hline\hline
			$d = 10, w = 16$ &     27.18      &     0.833      \\[1pt] \hline
			$d = 10, w = 32$ &    {27.70}     &    {0.848}     \\[1pt] \hline
			$d = 16, w = 32$ & \textbf{28.24} & \textbf{0.863} \\[1pt] \hline
			$d = 20, w = 16$ &     27.21      &     0.852      \\[1pt] \hline
			$d = 20, w = 32$ &     27.08      &     0.828      \\ \bottomrule[1.2 pt]
	\end{tabular}}
	\label{depthandwide}
\end{table}

\begin{table}
	\caption{Ablation studies on ELU, Bilinear, Wide, Dilated rate, and WRM. The PSNR Performance on Gopro test dataset.}
	\centering
	\resizebox{7.5cm}{!}{
		\begin{tabular}{|c| c c c c | c|}
			\hline
			Operation         & ELU       & Bilinear  & Dilated rate & WRM       & PSNR           \\[1pt] \hline
			\multirow{5}{*}{Baseline} & \ding{56} & \ding{56} & \ding{56}     & \ding{56} & 27.36          \\[1pt]
			& \ding{52} & \ding{56} & \ding{56}     & \ding{56} & 27.64          \\[1pt]
			& \ding{52} & \ding{52} & \ding{56}     & \ding{56} & 27.87          \\[1pt]
			& \ding{52} & \ding{52} & \ding{52}     & \ding{56} & {28.04}        \\[1pt]
			& \ding{52} & \ding{52} & \ding{52}     & \ding{52} & \textbf{28.36} \\[1pt] \hline
		\end{tabular}
	}
	\label{Ablation}
\end{table}

\begin{figure}
	\centering
	\includegraphics[scale=0.38,trim={2.5cm 5.8cm 5.5cm 3cm},clip]{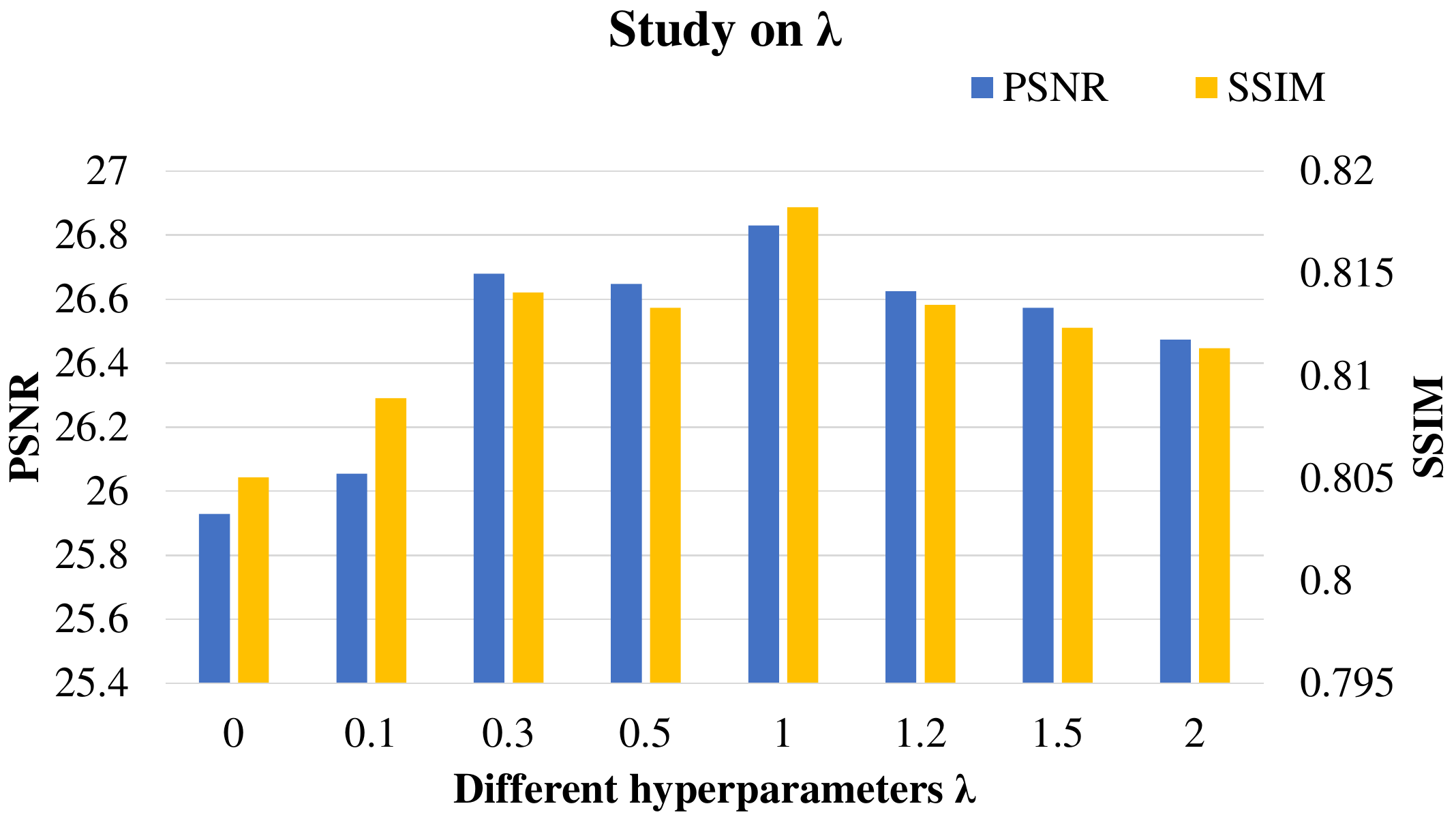}\\
	\caption{Study on the influence of $\lambda$. When $\lambda = 1$, the model achieves best results.}
	\label{loss}
\end{figure}

\subsection{Ablation Studies}
In this subsection, we design a series of ablation experiments to analyze the effectiveness of each of the modules we propose. We use the GoPro test set for evaluation and performed 200 epochs of training on an image patch of size $96\times96$. 

\textbf{Model Design Policy.} We explore the impact of different depths and widths on network performance, as shown in Table \ref{depthandwide}. Where the depth represents the number of DCBs we set and width represents the number of channels in our intermediate features. As can be seen from the experimental results, the width has a greater effect on our model than the depth. Our model works best at $d = 16$ and $w = 32$. Therefore, our final model is set to $d=16$ and $w=32$.

\textbf{Effectiveness of Each Operation.} We set up a baseline of DCBs, where the activation function is ReLU and the upsampling is deconvolution. We demonstrate the effectiveness of our proposed module by modifying the corresponding activation function and upsampling method, as shown in Table \ref{Ablation}. From the experiment, it is known that the ELU activation function and bilinear upsampling obtain better performance than ReLU and deconvolution. We adjust the dilated rate of the last layer of the DC module to help improve the performance of the network. And the WRM can help the network recover high-frequency details in the frequency domain, enabling network performance to be improved. These comparisons show that our proposed methods are useful for image deblurring.

\textbf{Effectiveness of $\lambda$ in Loss Function.} We conduct trade-off experiments for the Charbonnier loss and SSIM loss, as shown in Figure \ref{loss}. We find the optimal hyperparameter $\lambda$ by adjusting the value of the hyperparameter $\lambda$ so that the network performance can be optimized. We set the hyperparameters $\lambda$ to 0, 0.1, 0.3, 0.5, 1 and 2 respectively. We can notice from the graph that the best performance is obtained when the hyperparameter $\lambda = 1$.

\section{Conclusion}
In this work, we propose a novel dilated convolution network structure for image deblurring. For this structure, we propose two modules: the dilated convolution module and the wavelet reconstruction module. Specifically, the dilated convolution module use dilated convolution with different dilated rates, which effectively helps the network to obtain different receptive field information. The wavelet reconstruction module exploits the properties of the wavelet transform to provide high-frequency information for spatial domain reconstruction, resulting in clearer images. The quantitative and qualitative results show that our algorithm can effectively restore a clearer image than other methods. Our approach is simple in structure and easily transferable to other high-level tasks. In the future, this approach will be explored to facilitate other image restoration tasks such as image denoising and super-resolution.

\section*{Acknowledgments} 
This work was supported in part by the National Nature Science Foundation of China under Grant No. 61901117, U1805262, 61971165, in part by the Natural Science Foundation of Fujian Province under Grant No. 2019J05060, 201-9J01271, in part by the Fujian Provincial Education Department Project under Grant No. JT180094, JT180095, in part by the Special Fund for Marine Economic Development of Fujian Province under Grant No. ZHHY-2020-3, in part by the research program of Fujian Province under Grant No. 2018H6007, the Special Funds of the Central Government Guiding Local Science and Technology Development under Grant No. 2017L3009, and the National Key Research and Development Program of China under Grant No. 2016YFB-1001001.

{\small
	\bibliographystyle{unsrt}
	\bibliography{ref}
}

\end{document}